\documentstyle[12pt]{article}
\title{Vacuum expectations of the high dimensional operator and their
contribution in Bjorken and Ellis-Jaffe sum rules}
\vspace{2cm}
\author{A.G.Oganesian  \thanks{E-mail: armen@vitep5.itep.ru}\\
Institute of Theoretical and Experimental Physics,\\
RU-117259 Moscow, Russia}
\date{}
\begin{document}
\maketitle

\newcommand{\be}{\begin{equation}}
\newcommand{\ee}{\end{equation}}

\def\la{\mathrel{\mathpalette\fun <}}
\def\ga{\mathrel{\mathpalette\fun >}}
\def\fun#1#2{\lower3.6pt\vbox{\baselineskip0pt\lineskip.9pt
\ialign{$\mathsurround=0pt#1\hfil##\hfil$\crcr#2\crcr\sim\crcr}}}

\newpage

{\bf\large  Introduction}

\bigskip

It is well-known, that in some cases the contributions of the
high-dimensional ($d> 6$) operators in the QCD sum rules [1] are 
important. For estimation of the vacuum average of
high-dimensional operators usually is used factorization
hypothesis  [1]. According to this hypothesis it is
assumed that in the high-dimensional operators expansion over
intermediate states, vacuum states contributions are dominant.
Formally this assumption can be written as

$$< O_1 \cdot O_2 > \simeq < O_1> \cdot < O_2 >$$
where $< O >$  are some color scalar operators.

In [1] factorization hypothesis  was used to estimate vacuum
average of the 4-quarks operator $(d=6)$.
Gluons operators with $d=6$ and $d=8$ was been calculated in [2,3] (see 
also discussion about it in [4]). But for operators with  dimension
($d=7$) a large number of new vacuum averages appears, and some of
them can't be reduced to a product of vacuum averages of the operator with
lower dimension.  Nevertheless in this paper we make an attempt to estimate
vacuum averages of all operators with $d=7$. The method we offer is based on
factorization hypothesis.

It is well-known, that though factorization
hypothesis is confirmed in $1/N$ limit [1], but for real world
with N=3 factorization hypothesis in some cases has rather
bad accuracy (see for example [5]).
So in this work we shall use factorization hypothesis in
maximal "soft" form, i.e. we suppose:\\
1) If any operator can be saturated by vacuum intermediate states
(i.e. if vacuum intermediate states contribution exist
\underline{and is not zero}) we estimate it by its factorized value
and suppose accuracy 30\%.\\
2) To estimate other operators one should try
to express them through factorizable operators, if it is possible.\\
To avoid uncertainty, we also suppose:\\
3) Vacuum average of high-dimensional
operators should not depend on the way of factorization, and, particularly,
vacuum average of operators, containing derivatives should not depend on 
the fact are the equation of motion taken in account before or after
factorization.\\
This assumption 3) can be treated as a condition of
self-consistence of factorization hypothesis. This three assumptions we
hereinafter shall call anzatz.

The main idea is to consider the vacuum average of the
operator with large dimension $(d=10)$. Using fagtorization
hypothesis one can express it in terms of the product of
vacuum averages of operators with lower dimension $(d < 7)$,
some of which are known and some are unknown. According
condition of self-agreement of factorization one can write a
number of relations for this unknown operators and estimate
them. This method will be explicitly explained in section 1.
As by-product some estimations for vacuum average of the
operators of dimension 10 became available. Of course one
must note that this is only some rough phenomenological
estimations and do not claim for high accuracy. An accuracy
of our estimation are about a factor 2.

The paper is organized as follows.

 In section 1 we describe the method on an example of
calculation vacuum averages of the of dimension 7 operators,
constructed from quark and gluons.

In section 2 we discuss vacuum averages of the 7-dimension
operators with one derivatives. So it appears
possible to evaluate all vacuum averages
of the  dimension 7. In section 3 obtained results are
used to calculate high dimensional operator contribution to
Bjorken [6] and Ellis-Jaffe [7] sum rules.

\newpage
{\bf\large  Section 1}

\bigskip
In this section we'll discuss following vacuum averages of
the of dimension 7 nonfactorizable operators:

$$R_d = \frac{< g^2\bar{q}d^{nkl}\lambda^lG^n_{\mu\nu}G^k_{\mu\nu}q>}
{24};~~~
R_f = \frac{< g^2\bar{q}f^{nkl}\lambda^lG^n_{\mu\nu}G^k_{\mu\beta}
\sigma^{\nu\beta}q>}{24}$$

\be
S_1 = \frac{i< g^2\bar{q}\gamma_5  G^n_{\alpha\beta}
\tilde{G}^n_{\alpha\beta}\bar{q}>}{24};~~~
S_d = \frac{i< g^2\bar{q}\gamma_5 d^{nkl}\lambda^l
G^n_{\mu\nu}\tilde{G}^k_{\mu\nu}\bar{q}>}{24}
\ee
where $\tilde{G}^n_{\mu\nu}= G^n_{\mu_1\nu_1} \cdot
\varepsilon^{\mu\nu\mu_1\nu_1}/2$

For convenience hereinafter following notations will be used

$$\bar{R}_{d,f} = R_{d,f} < \bar{q}q>~~~~~~~~~\bar{R}_1 =
<g^2G^2> <\bar{q}q>^2/24$$

$$\bar{S}_{1,d} = S_{1,d} < \bar{q}q>~~~~~~~~~N =
(<g\bar{q} \hat{G}_{\mu\nu}\sigma^{\mu\nu}q>)^2/24 \eqno{\mbox{(1a)}}$$

$$\hat{G}_{\mu\nu} = \frac{\lambda^n}{2} G^n_{\mu\nu}; ~~~~~~~~~
G^2 = G^n_{\mu\nu}G^n_{\mu\nu}$$

Note, that at standard choose of gluon, quark and
quark-gluon condensates $\bar{R}\sim N$ .
We assume that in our accuracy vacuum averages for u and
d-quarks are the same, so, for example

$$< g^2 \bar{u}d^{nkl}\lambda^lG^n_{\mu\nu} G^k_{\mu\nu}u>=
<g^2 \bar{d}d^{nkl}\lambda^l G^n_{\mu\nu}G^k_{\mu\nu}d>$$
The method of estimation $R_d,~ R_f ...$       is based on our anzatz.
We consider some vacuum averages of the of dimension 10
operators and factorize them in two different ways. Once
we use equation of motion before factorization, and other
time -after. We require this two results to be the same
(with accuracy  $\sim 30$ \%, which is the accuracy of
factorization itself). Because of such uncertainty in order
to estimate vacuum averages (1) one have to use only those
equations, which are enough ($> 50$\%) sensitive to unknown
values of vacuum averages (1).

We shall illustrate the method on an example of the
following operator with dimension 10

$$< T_1 > = -<(\bar{d}[\nabla^2u]) \cdot (\bar{u}[\nabla^2d])>$$
where square of a bracket like $[\nabla^2u]$  mean that a
derivative acts only on the quark operator in bracket.
From one side $< T_1 >$ can be immediately factorized in the
single way

\be
< T_1 > = \frac{1}{12}<\bar{u}[\nabla^2 u]><\bar{d}[\nabla^2d]>
\ee
Using equations of motion  $\hat{\nabla}q=0$  and the fact that

\be
\nabla^2 = \hat{\nabla}\hat{\nabla} +\frac{1}{2}
g\hat{G}_{\mu\nu}\sigma^{\mu\nu}, ~~~~~~~(\mbox{where} ~\hat{\nabla} =
\gamma_{\mu}\nabla_{\mu})
\ee
One can find

\be
<T_1> = \frac{1}{2}\cdot N
\ee

where N is denoted in (1a).\\
From other side if take into consideration equation of
mouton from very beginning, one get

\be
< T_1 > =  -\frac{g^2}{4} < \bar{d}\frac{\lambda^k}{2}\sigma^{\alpha \beta}u
\cdot \bar{u}\frac{\lambda^n}{2}\sigma^{\mu\nu}d \cdot
G^n_{\alpha\beta}G^k_{\mu\nu}>
\ee
Now one can do here Fiertz transformation (for a simplicity we will write
down it only for color indexes, and for Lorenz indexes it is meant)

$$
< T_1 > = \frac{g^2}{4}\Biggl \{
\frac{1}{3}<(\bar{u}_{\tau^{\prime}} u_{\tau})\cdot
(\bar{d}_{\rho^{\prime}}\lambda^k \lambda^n
d_{\rho})G^k_{\alpha\beta}G^n_{\varphi\varepsilon}> +$$

\be
+ \frac{1}{2} <(\bar{u}_{\tau^{\prime}} \lambda^lu_{\tau})\cdot
(\bar{d}_{\rho^{\prime}} \lambda^k\lambda^l\lambda^nd_{\rho})
G^k_{\alpha\beta}G^n_
{\varphi\varepsilon} > \Biggr \} \cdot \frac{1}{4}\cdot \sigma^{\alpha\beta}_
{\rho^{\prime}\tau}\sigma^{\varphi\varepsilon}_{\tau^{\prime}\rho}
\ee
Using the fact that

$$
\Biggl ( \lambda^k\lambda^l\lambda^n\Biggr )^{ab} = \frac{2}{3}
\Biggl ( d^{nkl} + i f^{nkl}\Biggr ) \delta^{ab} + \frac{13}{21}
\Biggl ( (\lambda^k)^{ab}\delta_{ln} + (\lambda^n)^{ab}\delta_{lk} \Biggr )
- $$

\be
- \frac{5}{21}(\lambda^l)^{ab} \delta^{kn} + O^{mkln}(\lambda^m)^{ab}
\ee
(where $  O^{mkln} = 
Tr(\lambda^m\lambda^k\lambda^l\lambda^n)/2 - \frac{13}{21}
(\delta^{mk}\delta^{ln} + \delta^{mn}\delta^{kl} - \frac{5}{13}
\delta^{ml}\delta^{kn})$
- is traceless matrixes by each pair of
indexes)
we find

$$<T_1> = \frac{g^2}{16}\sigma^{\alpha\beta}_{\rho^{\prime}\tau}
\sigma^{\varphi\varepsilon}_{\tau^{\prime}\rho}\cdot \Biggl \{ \frac{1}{3}
<(\bar{u}_{\tau^{\prime}}u_{\tau})
\cdot(\bar{d}_{\rho_{\prime}}\lambda^k\lambda^nd_{\rho}
G^k_{\alpha\beta}G^n_{\varphi\varepsilon})> $$

$$+ \frac{1}{3} <(\bar{u}_{\tau^{\prime}}\lambda^l(d^{nkl} + if^{nkl})
G^k_{\alpha\beta}G^n_{\varphi\varepsilon}u_{\tau})\cdot
(\bar{d}_{\rho^{\prime}}d_{\rho})> $$

$$ + \frac{26}{21}\Biggl [
<(\bar{u}_{\tau^{\prime}}\hat{G}_{\varphi\varepsilon} u_{\tau})\cdot
(\bar{d}_{\rho^{\prime}}\hat{G}_{\alpha\beta}d_{\rho})> +
<(\bar{u}_{\tau^{\prime}}\hat{G}_{\alpha\beta}u_{\tau})
(\bar{d}_{\rho^{\prime}}\hat{G}_{\varphi\varepsilon}d_{\rho})>\Biggr ]$$

$$- \frac{5}{42} <(\bar{u}_{\tau^{\prime}}\lambda^l
u_{\tau})(\bar{d}_{\rho^{\prime}}\lambda^l
d_{\rho})G^k_{\alpha\beta}G^n_{\varphi\varepsilon}>$$

\be
+ \frac{1}{2}<(\bar{u}_{\tau^{\prime}}\lambda^l u_{\tau})
(d_{\rho^{\prime}}O^{mkln}\lambda^m
d_{\rho})G^k_{\alpha\beta}G^n_{\varphi\varepsilon}> \Biggr \}
\ee
Fourth term in (8) after factorization appears to be expressed in terms
$<(\bar{u}\lambda^l\Gamma u)\cdot (\bar{d}\lambda^l\Gamma d)><g^2 G^2>$
where $\Gamma=1, ~\gamma_5,~\gamma_{\mu},
~\gamma_5\gamma_{\mu},~\sigma^{\mu v}$. This terms assumed to be $0$,
because, as was shown in [5], $<(\bar{u}\lambda^l\Gamma
u)\cdot(\bar{d}\lambda^l\Gamma d)>$ are negligible small for
$\Gamma=1,\gamma^{\mu}$ and there is not any reasons to believe that for
other $\Gamma$ result will grow on an order, so, we can neglect by fourth
term. The last, fifth, term in (8) appears to be  $0$
at factorization due to fact that $O^{nklm}$ is traceless matrixes
The first three terms in (8) allows vacuum intermediate state, so $< T_1 >$
can be factorized and after some calculations one can found

\be
< T_1 > = \bar{R}_1/6 + \bar{R}_d/2 + \bar{R}_f + \frac{13}{14}N
\ee
Here we omit terms, proportional to $\bar{S}_1,~\bar{S}_d$, because, as
will be shown later, they are negligibly small.  Comparing (4) and (9) we
get

\be
N/2 = \bar{R}_1/6 + \bar{R}_d/2 + \bar{R}_f + \frac{13}{14}N
\ee
Note that this equality is rather sensitive to unknown vacuum
averages $\bar{R}_d$   and  $\bar{R}_s$. Really, if we suppose
$\bar{R}_d=\bar{R}_f=0$, then left
and right side differ from each other more than twice.

 One more equation can be obtained, if we consider such an operator

\be
< T_2 > =  <\bar{d}_{\lambda}[L^{\alpha\mu}u]_{\rho}) \cdot
(\bar{u}_{\tau^{\prime}} u_{\tau})> \cdot \delta^{\tau^{\prime}\tau} \cdot
(\gamma^{\alpha}\gamma^{\mu})_{\lambda\rho}
\ee
where $L^{\alpha\mu}=(\nabla_{\alpha}\nabla_{\beta}
\nabla_{\mu}\nabla_{\beta} - \nabla_{\beta}\nabla_
{\alpha}\nabla_{\mu}\nabla_{\beta})$.\\
$< T_2 >$ can be factorized immediately as $< T_2 > =
-<\bar{u}\gamma^{\alpha}\gamma^{\mu}[\prod^{\alpha\mu} u]><\bar{d}d> \cdot
\frac{1}{12}$.\\
Using equations of motion after some calculations one can find

\be
< T_2 > = \bar{R}_1/3 + \bar{R}_d/2 + \bar{R}_f/2
\ee
From other side by use of equation of motion one can write
(11) in the form:\\

$$< T_2 > = -g^2<(\bar{d}\hat{G}_{\alpha\beta}
\hat{G}_{\mu\beta}\gamma^{\alpha}\gamma^{\mu}u)(\bar{u}d)>$$
Then
after Fiertz transformation just as in previous case one can
get

\be
< T_2 > = \bar{R}_1/3 + \bar{R}_d + \bar{R}_f + \frac{11}{21}N
\ee
From (12) and (13) we have

\be
\bar{R}_1/3 + (\bar{R}_f + \bar{R}_d)/2 = \frac{\bar{R}_1}{3} + \bar{R}_d +
\bar{R}_f + \frac{11}{21}N
\ee
Note, that (14) also is sensitive to
$\bar{R}_d$, $\bar{R}_f$.\\ Using the same procedure for the operator

\be
< W > = -<i(\bar{d}\gamma_5\varepsilon^{\alpha\beta\mu\nu}[\nabla_{\alpha}
\nabla_{\beta}\nabla_{\mu}\nabla_{\nu}u])(\bar{u}d)>
\ee
we get such an equation:

\be
\bar{S}_1/3 + \bar{S}_d/2 = \bar{S}_1/3 + \bar{S}_d - \frac{2}{21} N
\ee
From (16) we can conclude $\bar{S}_d \sim N/5 \ll N$ so we put it 0. What
about $\bar{S}_1$, we assume it to be negligibly small too.
Really, it is easy to show by direct calculations

\be
S_1 = \frac{1}{24} < \bar{q}q \cdot G^2 > -
\frac{1}{48}<\bar{q}\sigma^{\alpha\beta}\sigma^{\mu\nu}~
G^n_{\alpha\beta}G^n_{\mu\nu} q >
\ee
So $S_1$  is equal to difference of two factorizable vacuum
averages which cancel each other after factorization. So we
can expect that $S_1$ should be much more less then this vacuum
averages (which are of order of $R_1$) and we can put it 0
within our accuracy.

From (10), (14) we found

$$\bar{R}_f \sim 4/21 \cdot N - \bar{R}_1/3 \ll \bar{R}_1~~~(\mbox{or} ~~~
N); ~~~ \mbox{so}~~\bar{R}_f \sim 0$$

$$\bar{R}_d \sim -26/21 \cdot N + \bar{R}_1/3 \sim -N~~~
(\mbox{or}~~~\bar{R}_d\sim  -\bar{R}_1~~~\mbox{because}~~~
\bar{R}_1 \sim N, ~~\mbox{see~~~(1)})$$

   Finally, we write down results of this section:

\be
R_d \sim -R_1, ~~~S_1 \sim S_d \sim R_f \approx 0
\ee

\bigskip

{\bf \large Section 2}

\bigskip
In this section we will discuss vacuum averages of the
four-quark operators with one derivative, such us:

$$X_1 = <(\bar{q}[\nabla_{\alpha}q])(\bar{q}\gamma^{\alpha}q)>;~~~
\bar{X_1} =
<(\bar{q}\gamma_5[\nabla_{\alpha}q])(\bar{q}\gamma_5\gamma^{\alpha}
q)>$$

$$X_2 =
<(\bar{q}\lambda^k[\nabla_{\alpha}q])
(\bar{q}\lambda^k\gamma^{\alpha}
q)>;~~~
\bar{X_2} =
<(\bar{q}\lambda^k\gamma_5[\nabla_{\alpha}q])
(\bar{q}\lambda^k\gamma_5\gamma^{\alpha}
q)>$$

$$Y_1 = i <
(\bar{q}\gamma^{\tau}[\nabla_{\varepsilon}q])(\bar{q}\sigma^
{\tau\varepsilon}q) >;~~~
\bar{Y_1} = i <
(\bar{q}\gamma_5\gamma^{\tau}[\nabla_{\varepsilon}q])
(\bar{q}\gamma_5\sigma^{\tau\varepsilon}q) >$$

\be
Y_2 = i <
(\bar{q}\lambda^k\gamma^{\tau}[\nabla_{\varepsilon}q])
(\bar{q}\lambda^k\sigma^{\tau\varepsilon}q) >;~~~
\bar{Y}_2 = i <
(\bar{q}\lambda^k\gamma_5\gamma^{\tau}[\nabla_{\varepsilon}q])
(\bar{q}\lambda^k\gamma_5\sigma^{\tau\varepsilon}q) >
\ee
Note, that one can't factorize this operators immediately (after proper
Fiertz transformation), because due to equation of motion they became 
zero after
factorization (see point 1,2 of our anzatz).
Note also, that all other vacuum averages of the four-quark operators with
one derivative easily can be expressed by this eight, by help of
equations of motions.
   Of course, this eight operators aren't independent.
First it can easily be shown, that $Y_1=-X_1$. Really, due to C-parity

\be
Y_1 = \frac{i}{2}<(\bar{q}\gamma^{\tau}[\nabla_{\varepsilon}q] +
[\bar{q}\nabla_{\varepsilon}]\gamma^{\tau}q) \cdot
(\bar{q}\sigma^{\tau\varepsilon}q)> = \frac{i}{2}<
[\partial_{\varepsilon}(\bar{q}\gamma^{\tau}q)] \cdot
(\bar{q}\sigma^{\tau\varepsilon}q)>
\ee
Neglecting the full derivatives (and all possible anomalies) one can write

$$Y_1 = -\frac{i}{2}<(\bar{q}\gamma^{\tau}q)\cdot
([\bar{q}\nabla_{\varepsilon}]\sigma^{\tau\varepsilon}q +
\bar{q}\sigma^{\tau\varepsilon}[\nabla_{\varepsilon}q])> $$
Now, using equations of
mouton $\hat{\nabla}q = 0$ it is easy to find

\be
Y_1 = -\frac{i}{2}<(\bar{q}\gamma^{\tau}q)\cdot ([\bar{q}\nabla_{\tau}]q -
\bar{q}[\nabla_{\tau})q])>
\ee
Finally from C-parity it is clear that

\be
Y_1 = -X_1
\ee
In the same way, if we neglect anomalies, one can show


$$\bar{X}_1 = -<(\bar{q}\gamma_5\gamma^{\varepsilon}q)\cdot
\frac{1}{2}([\bar{q}\nabla_{\varepsilon}]\gamma_5q + 
\bar{q}\gamma_5[\nabla_{\varepsilon}q])> =$$
\be
=-\frac{1}{2}<([\bar{q}\nabla_{\varepsilon}]\gamma_5\gamma^{\varepsilon}q + 
\bar{q}\gamma_5\gamma^{\varepsilon}[\nabla_{\varepsilon})q])
({\bar{q}\gamma_5q})> = 0
\ee
Using Fiertz transformation both by color and scalar indexes, one can express
vacuum averages (19) through each other. Finally a system of exact
equations can be found, which solution, taking in account (22,23),  is:

\be
Y_2 = -\bar{Y}_2 = -X_2 = \frac{8}{3} X_1;~~~Y_1 = -\bar{Y}_1 = -X_1; ~~~
\bar{X}_1 = \bar{X}_2 = 0
\ee
So we see, that all vacuum averages of operators, constructed from
four quarks and one derivative are expressed through  $X_1$.
One must emphasize, that (24) is exact statement, don't
based on factorization  hypothesis.

   To estimate $X_1$, we use factorization hypothesis analogously
to what we have done in \\
sect. 1. Let us consider vacuum averages
of the operator $Z_1$

\be
Z_1 = <2(\bar{q}\lambda^n(D_{\alpha}G_{\alpha\beta})^nq) \cdot (\bar{q}
[\nabla_{\beta}q]) >
\ee
which can be factorized as

\be
Z_1 = -
\frac{1}{6}<(\bar{q}\lambda^n(D_{\alpha}G_{\alpha\beta})^n[\nabla
_{\beta}q])>
<\bar{q}q>
\ee
Now we can use equations of motion
$D_{\alpha}G^n_{\alpha\beta}=-g\bar{q}(\lambda^n/2)\gamma^{\beta}$
(for simplicity we
shall limit us by case with one flavor, two flavor case is similar). Then,
taking into account (24), we get:

\be
Z_1 =
\frac{g}{12}<(\bar{q}\lambda^n[\nabla_{\beta}q])(q\lambda^n\gamma^
{\beta}q)><\bar{q}q> = \frac{g}{12}X_2 <\bar{q}q> = -\frac{2}{9}X_1
<\bar{q}q> \cdot g
\ee
On the other hand, using equations of motions, (25) can be rewritten as

\be
Z_1 =
-<(\bar{q}\lambda^nq)(\bar{q}\gamma^{\beta}\lambda^nq)(\bar{q}[\nabla_
{\beta}q])>\cdot g
\ee
Now, using Fiertz transformation and also use C-parity and neglecting
full derivatives (in the same way as in (21-22)), we can write

\be
Z_1 = \frac{1}{2}\cdot \Biggl \{
\frac{7}{3}<(\bar{q}q)(\bar{q}\gamma^{\beta}q) (\bar{q}[\nabla_{\beta}
q])> + < (\bar{q}\gamma_5\gamma^{\alpha}
\varepsilon^{\alpha\beta\tau\varepsilon}q) \cdot
(\bar{q}\sigma^{\tau\varepsilon}q) \cdot (\bar{q}[\nabla_{\beta}q])>
\Biggr\}
\ee
The first term here can be factorized, so according our anzatz we
have

\be
Z_1 = \frac{7}{6}<\bar{q}q> \cdot X_1 \cdot g
\ee
Comparing (27) and (30) we get $\frac{7}{6}~X_1 = -\frac{2}{9}~X_1$,
So

\be
X_1 = 0
\ee
Then from (24) we can conclude, that all vacuum averages (19) are zero.
One can easily see, that every vacuum averages of the
operators with dimension 7 can be expressed through a set
vacuum averages, discussed in this two sections. Thus,
results, obtained in sections 1,2 allows one to give estimations
for all possible vacuum averages of the operators with
dimension 7.
 This estimation may be significant in a large range of
problem, where contribution of high dimension operators
became necessary.

An example of such problem is the calculation of Bjorken [6] and
Ellis -Jaffe [7] sum rules in the framework of QCD sum rules,
offered in [8] we are going to discuss in next section

\bigskip

{\bf \large Section 3}

\bigskip
In this section we use results obtained in the previous
 sections for the analysis of the power corrections to the
first moment of the structure function of a polarized nucleon.
On importance of the power corrections $1/Q^2$  to structure
functions of polarized nucleon was
indicated in [9,10], where necessity of their contribution was
considered to satisfy with experimental data on deep
inelastic scattering on a polarized [11] nucleon, on the one
hand, and Gerasimov - Drell - Hearn sum rule [12,13] - with other
(see also discussion
of this problem in connection  with "spin-crisis" problem in review [14,15]).
By the most natural candidate for this role seems the contribution of the
operators of twist 4.  The contribution of these operators to deep inelastic
scattering on a polarized nucleon was calculated in [16].

$$M^{S(NS)}\equiv \int dx~g_1^{p+n,(p-n)}(x,Q^2)=$$

$$= K^{S(NS)}\cdot \Biggl \{ g_A^{S(NS)}(1 - \alpha_s(Q^2)/\pi) -
\frac{8}{9} \frac{\ll U^{S(NS)}\gg}{Q^2}\Biggr \}+$$

\be
+ \frac{4}{3}\frac{m^2_N}{Q^2}\int dxx^2 \Biggl (g^{p+n,(p-n)}_2 (x)
 + \frac{5}{6} g_1^{p+n,(p-n)}(x) \Biggr ) + O(1/q^4)
\ee
Here:

$$K^{S(NS)} = \frac{5}{18}(\frac{1}{6}); ~~~~ g^{NS}_A = \left |
\frac{G_A}{G_V}\right | = 1.25; ~~~~ g^S_A = 0.1 \pm 0.04
~~~(\mbox{see}~[14])$$
$\ll U^{S(NS)}\gg$ are defined as $< N\mid U_{\mu}^{S(NS)}\mid N >$
$ = S_{\mu}\ll U^{S(NS)}\gg$

$$U^S = \bar{u}~\hat{\tilde{G}}_{\mu\nu}\gamma_{\nu}u + (u\to d) +
\frac{18}{5}(u\to s)$$

$$U^{NS} = \bar{u}~\hat{\tilde{G}}_{\mu\nu}\gamma_{\nu}u - (u \to d)$$
and $S_{\mu} = \bar{N}\gamma_{\mu}\gamma_5 N$; $N$ be a nucleon spinor. $\ll
U^{S(NS)}\gg$ are matrix element of twist four, we are interest in this
paper.
$\ll U \gg$ has been calculated in [8] from sum rules for 3-point
correlator

\be
\Gamma_{\mu}(p) = i^2 \int dx e^{ipx}\int dy < T
[\eta(x)U^{S(NS)}_{\mu}(y)\bar{\eta}(0)]> = -2p_{\mu}\hat{p}\gamma_5
\frac{\lambda^2_p\ll U^{S(NS)} \gg}{(m^2_N - p^2)^2} + ...
\ee
Where $\lambda_p$ is proton coupling and $\eta$ is proton current [17]

$$\eta = \varepsilon^{abc}(u^a
C\gamma_{\lambda}u^b)\gamma_5\gamma_{\lambda}d^c$$
As always in QCD sum rules, the correlator (33) is considered at
large negative $p^2$. However in this case, in difference
from usual 3-point correlator, though  $x \sim 1/p$ is small,
but there are no limitations on y and, therefore, it is
necessary to take into account region   $y\gg x \sim 1/p$ too. This
lead to the fact, that except usual vacuum expectations (local operators) in
operator expansion also appears field induced vacuum expectations - 2-point
(bilocal) correlators of the type

$$i~ \int d^4y < T \{ O_{\mu}(y)O^{\mu}(0)\} >$$
(see papers [18 - 21], where this
approach was offered and discussed).

In some cases these bilocal operators can be reduced to
local ones, using low -energy relations (see, for example
[18,22]), in other cases one should consider corresponding
2-point sum rules to estimate this bilocal operators (see
[20,21], for example).

In the [8] the correlator (33) was calculated and the power
correction in $1/p^2$ up to contribution of the operators
with dimension $d=8$ was accounted. After borelization the result, obtained
in [8], has the form

$$\ll U \gg + R \cdot M^2 = - \frac{exp(m^2_N/M^2)}{2\lambda^2_p}
\Biggl ( 2AM^2 \int\limits^{s_0}_0 ds~s^2 e^{-s/M^2}ln~\mu^2/s +$$

\be
+ BM^4(1 - exp(-s_0/M^2)) + CM^2 + D \Biggr )
\ee
A, B, C, D correspond to loop contribution (A) and power correction of
operators of dimension 4,6,8\footnote{For the
scalar case $\ll U^S \gg$ contribution of $s$-quark in [8] was neglected.}.
(Here, and also in expressions (35, 36,40) we for simplification 
omit subscript S(NS)  in all cases, where it is obvious). 
Note, that (34) depend on ultraviolet cut-off parameter $\mu^2$, even after
borelization. This is consequence of a simple, but incorrect model of
continuum accounting in [8], based on ordinary dispersion relation, as was
noted in [14]. In [14] was offered the method, how one should
correctly exclude continuum, using double dispersion relations and was shown,
that in this method dependence of unphysical cut-off parameter $\mu^2$
disappear. The procedure, offered in [14] lead to following sum rules
(instead of (34))

$$\ll U \gg + R \cdot M^2 = - \frac{exp(m^2_N/M^2)}{2\bar{\lambda}^2_p}
\Biggl \{ 2AM^2 \int\limits^{s_0}_0 ds~ s \cdot$$

\be
\cdot e^{-s/M^2}(s_0 + s~ln~s_0/s) + BM^4 \Biggl (1-(1 +
\frac{s_0}{M^2})e^{-s_0/M^2} \Biggr ) + CM^2 + D \Biggr \}
\ee
Here:

$$A^{S} = \alpha_s/\pi \cdot 4/5;~ B^{S} = -32/3 \cdot \pi^2
f^2_{\pi}\delta^2$$

$$A^{NS} = \alpha_s/\pi \cdot 4/9;~ B^{NS} = 
 -\frac{< g^2G^2>}{9} $$

$$C^{S} = (ln ~s_0/M^2 + 0.5) \cdot 32/27 \cdot \alpha_s/\pi \cdot
a^2 + 8/9 \cdot \pi^2 \cdot \Pi$$

$$C^{NS} = (ln ~s_0/M^2 + 1) \cdot 32/27 \cdot \alpha_s/\pi \cdot
a^2 + 8/9 \cdot \pi^2 \cdot \Pi$$

$$D^{S} = -1/9 \cdot m^2_0 \cdot 2 \cdot a^2$$

$$D^{NS} = -1/3 \cdot m^2_0 \cdot 2 \cdot a^2$$

$$a = -4\pi^2 <\bar{\psi}\psi>$$

$R$ correspond to the contribution of single-pole terms, $\Pi$ 
is bilocal power correction, which was estimated in [8] 
as $\Pi=3.10^{-3}$GeV$^6$ ;
 other parameters are standard:
 $f_{\pi} = 0.133$GeV, ${\delta}^2 = 0.2$ GeV$^2$ [23], $m^2_0 \simeq
0.8$GeV$^2$ [17]; $\alpha_s(1$GeV)$\sim0.37,~ <\bar{\psi\psi}>=-0.014$GeV$^3$,
$<g^2G^2>=0.5$GeV$^2~$;
$\bar{\lambda}^2_p = 32 \pi^4 \lambda^2_p = 2.1$ GeV$^6$, 
and continuum threshold $s_0 = 2.25$GeV$^2$ ([17, 21], see also [14]).
Hereafter we will use this correct result for $\ll U \gg$ (35), but one
should note, that results and conclusions we will present at the end of this
section, are similar both for (34) and (35).\\

To single out
$<< U> >$ itself, according to [8] the operator

$$1 - M^2 \frac{d}{dM^2}$$
acting on both sides of (34) was
used.  Using this operator for improved result (35), we find

$$\ll U \gg = \frac{-e^{m^2_N/M^2}}{2\bar{\lambda}^2_p}\Biggl [
2A\int\limits^{s_0}_0 ds~s(m^2_N - s)e^{-s/M^2} (s_0 + s~ln~s_0/s) + $$

$$
+ B\Biggl ( (m^2_N - M^2)(1 - (1 + \frac{s_0}{m^2})e^{-s_0/M^2}) +
\Biggl (\frac{s_0}{M^2}\Biggr )^2 ~e^{-s_0/M^2} \Biggr ) +$$

\be
+ (C m^2_N + 32/27 \cdot \alpha_s/\pi \cdot a^2 \cdot M^2)
+ D(1 + m^2_N/M^2)\Biggr ]
\ee

Analysis of sum rules (34), according [8], lead to
$ \ll U^S \gg \simeq -(0 \div 0.1) \mbox{GeV}^2$,
$ \ll U^{NS} \gg \simeq 0.18$  at Borel mass $M^2 \simeq 1$GeV$^2$
and even we take into account improved form (35,36), results change slightly
(see [14]). However from (34-36) it is possible to see, that in $\ll U \gg$
just dominates the contribution of the operators of dimension 8,
(both for (34) or (35,36)) that is the
last accounted term of expansion. Thus, there is the problem on a
reliability of results, obtained in [8], and for this purpose it is
necessary to evaluate the following term in expansion, that is contribution
of the operators of dimension $D=10$. As usual, we shall consider that sum
rules (36) are reliable, if contribution of the operators of dimension
$D=10$ will appear less than contribution of the operators of dimension
$D=8$.

In this section we will estimate the dimension 10 operators
contribution to sum rules for $\ll U\gg$ (36). We will take into
account only tree diagrams (fig.1a-1d), because all other are
suppressed by loop factors like  $1/4\pi^2$.

The main problem here is the estimation of vacuum averages
of dimension 10 operators, and this can be done by help of
results of sect.1,2. In this section we'll calculate the
contribution off all local operators of D=10 (fig.1a,b) and
also those bilocal operators (fig.1c), for which exact
low-energy relations exists (see eq.(27) in [17], where this
relation was discussed for very similar case).

In this work we don't take in account diagrams of fig.1d,
because they consist bilocal operator contribution, the
calculation of which is a separate problem. Nevertheless the
estimations allow to hope, that the contribution of this
diagram will hardly essentially change an obtained result.

It is necessary to note, that in [17] similar 3-point
correlator for the current
$\bar{q}\hat{\tilde{G}}_{\mu\nu}\gamma_{\nu}\gamma_5q$ was considered up to
dimension 10. But in our case we take into account greater number
of the diagrams because, using results of the previous sections and our
anzatz, we can evaluate practically all arising vacuum averages of dimension
10, and not just only those from them, which can immediately factorized
to forms $< g^2G^2 ><\bar{\psi}\psi>^2$ or
$(<\bar{\psi}\hat{G}_{\mu\nu}\sigma_{\mu\nu}\psi>)^2$.

Let us also do some notices about the diagram on fig.1b. On
the first sight it express in terms of unfactorizible vacuum
average like:

\be
K = < \bar{u}_{\lambda}g\hat{\tilde{G}}_{\mu\nu}
[\nabla_{\alpha}u_{\rho}] \cdot \bar{u}_{\tau}
[\nabla_{\beta}u_{\sigma}]> \cdot
T^{\mu\nu\alpha\beta}_{\lambda\rho\tau\sigma}
\ee
where $T^{\mu\nu\alpha\beta}_{\alpha\rho\tau\sigma}$ - any matrix,
constructed from $\gamma^{\nu}$-martrixes (and
$g^{\mu\nu},$ $\varepsilon^{\mu\nu\alpha\beta}$ also). However up to full
derivatives (37) can be written as

$$ K = -T^{\mu\nu\alpha\beta}_{\lambda\rho\tau\sigma}\Biggl \{
< (
[\bar{u}_{\lambda}\nabla_{\beta}]\hat{\tilde{G}}_{\mu\nu}[\nabla_{\alpha}
u_{\rho}]) \cdot (\bar{u}_{\tau}u_{\sigma})> +$$

$$+
<(\bar{u}_{\lambda}(D_{\beta}\hat{\tilde{G}}_{\mu\nu})[\nabla_{\alpha}
u_{\rho}])\cdot (\bar{u}_{\tau}u_{\sigma})>$$

$$+ <(\bar{u}_{\lambda}\hat{\tilde{G}}_{\mu\nu}[\nabla_{\beta}
\nabla_{\alpha}u_{\rho}]) \cdot (\bar{u}_{\tau}u_{\sigma}) > + $$

\be
+ <(\bar{u}\hat{\tilde{G}}_{\mu\nu}[\nabla_{\alpha}u_{\rho}])
([\bar{u}_{\tau}\nabla_{\beta}]u_{\sigma}) >\Biggr \}
\ee
Now one can easily
see that all vacuum averages in right side of (37) can be factorized.

All other diagrams (fig.1a, 1b) can be immediately
factorized by use of results (18),(24),(31).  Finally, result for sum rules
(36), taking into account the contribution of operators of dimension 10
 (fig.1a-1c), is:

$$\ll U \gg = \frac{-e^{m^2_N/M^2}}{2\bar{\lambda}^2_p}\Biggl [
2A\int\limits^{s_0}_0 ds~s(m^2_N - s)e^{-s/M^2} (s_0 + s~ln~s_0/s) + $$

$$
+ B\Biggl ( (m^2_N - M^2)(1 - (1 + \frac{s_0}{m^2})e^{-s_0/M^2}) +
\Biggl (\frac{s_0}{M^2}\Biggr )^2 ~e^{-s_0/M^2} \Biggr ) +$$

\be
+ (C m^2_N + 32/27 \cdot \alpha_s/\pi \cdot a^2 \cdot M^2)+D(1 + m^2_N
/M^2)+\frac{E}{M^2}\Biggl (1 + \frac{m^2_N}{2m^2} \Biggr )\Biggr ]
\ee
where contribution of dimension 10 operators  are

$$E^{NS} = \frac{<\bar{\psi}\psi>^2}{18} \Biggl \{ m^4_0 -  \Biggl
(\frac{10}{9}<g^2G^2> + \frac{11}{6} m^4_0\Biggr )\Biggr \} \simeq
-\frac{1}{9}m^4_0 <\bar{\psi}\psi>^2$$

\be
E^S = \frac{<\bar{\psi}\psi>^2}{18} \Biggl \{ m^4_0 + \Biggl
(\frac{10}{9}<g^2G^2> + \frac{11}{6} m^4_0\Biggr )\Biggr \} \simeq
\frac{2}{9}m^4_0 <\bar{\psi}\psi>^2
\ee
In (40) we omit all vacuum averages
like $\bar{R}_f$, or $\bar{S}_1$ or
$X_1\cdot <\bar{\psi}\psi >$ and so on, which have been shown in sect.1,2 to
be negligible small, and also take into account result for $\bar{R}_d$ from
(18).

Using standard values $< g^2G^2 >, <\bar{\psi}\psi >$  one can see, that
for $\ll U^{NS}\gg $  contribution of dimension 10 is only about
20-25 \% of those of dimension 8 and are within the limits of
permissible accuracy. So for $\ll U^{NS}\gg $  our results confirm a
conclusion, made in [8], that contribution of operators of
twist 4 in Bjorken sum rules (6) are small.

But for case $\ll U^S \gg$ contribution of dimension 10 became
approximately more or equal  that contribution of
dimension 8, (at $M^2=1GeV^2$)  so for $\ll U^S \gg$ it seems that sum
 rules (35,39) are
 inapplicable. So, most likely, for twist 4 operators
contribution to $M^S$ in (32) (i.e. in for Ellis-Jaffe [7] sum rules)
it is impossible
to make any predictions from QCD sum rules (at least in those approach as
discussed). Note also, that some other real reasons, that also make result
for $\ll U^S \gg$ doubtful are discussed in [14].

Author would like to thank B.L.Ioffe for
very precious discussions and many useful advises.

This work is supported in part by CRDF grant RP2-132, INTAS 93-0283
and Schweizerishe National Fond 7SUPJ048716

\newpage

\centerline{\bf \large  References}

\bigskip
\noindent
1. M.A. Shifman., A.I. Vainstein, V.G. Zakharov,
   Nucl. Phys.B147 (1979)\\
2. S. N. Nikolaev, A.V. Radjushkin,
      Nucl. Phys.B213 (1983), 285.;
   Phys. Lett.B110 (1982) 476. \\
3. S. N. Nikolaev, A.V. Radjushkin,
   Phys. Lett.B124 (1983) 243.\\
4. V.A. Novikov et al., Nucl. Phys.B237 (1984), 525.\\
5. A.R. Zitnitsky, Yad. Fiz.41 (1985),  805, 1035, 1331.\\
6. J.D. Bjorken, Phys. Rev 148 (1966), 1467. \\
7. J. Ellis, R.L. Jaffe, Phys. Rev.D9 (1974), 1444; D10 (1974), 1669(E).\\
8. I.I. Balitsky, V.M. Braun, A.V. Kolesnichenko,
    Phys. Lett.B242 (1990), 245.; B318 (1993), 648. (E)\\
9. M. Anselmino, B.L. Ioffe, E. Leader, Yad. Fiz. 49 (1989), 214.\\
10. V.D. Burkhert, B.L. Ioffe, ZhETF 105,(1994) 1153.\\
11. J. Ashman et al., Nucl.Phys.B238 (1989) 1.\\
12. S. B. Gerasimov Yad. Fiz.2 (1965), 598\\
13. S.D. Drell, A.C. Hearn Phys.Rev.Lett. 16 (1966) 908\\
14. B.L.Ioffe  Preprint ITEP  62-95, 1995; Yad. Fiz. 58 (1995), 1492.\\
15. B.L. Ioffe, Surveys in High Energy  Physics, 1995, Vol.8. 107.\\
16. E.V. Shuryak, A.I. Vainstein,  Nucl. Phys.B 201 (1982), 144;

    X. Ji, M. Unrau, Phys.Lett B333 (1994) 228\\
17. B.L. Ioffe, Nucl. Phys.B 188 (1981) 317, B191 (1981) 591. (E).

    V.M.Belyaev, B.L.Ioffe, Sov.Phys.JETP 56 (1982) 493.\\
18. V.M. Braun, A.V. Kolesnichenko, Nucl. Phys.B283 (1987), 723.\\
19. I.I. Balitsky, A.V.Yung Phys. Lett.B 129, (1983) 328.\\
20. V. M. Belyaev, B.L. Ioffe, Ya. I. Kogan, Phys.
Lett.B151 (1985), 290.\\
21. B.L. Ioffe, A.V.Smilga, Nucl. Phys.B 232 (1984), 109\\
22. V.M. Braun, A.V. Kolesnichenko, Yad. Fiz. 44 (1986) 756.\\
23. V.A.Novikov et al Phys.Lett. B86 (1979) 347
\newpage
\centerline{\bf \large Figure captions}

\bigskip
\noindent
Fig. 1. Diagrams, corresponding of the  dim.10 contribution.
Circles denote derivatives,

~~~dashed lines denote gluons, solid lines
denote quarks.
\end{document}